\documentclass[12pt]{article}
\usepackage{amsmath,amssymb,epsfig,color}
\usepackage[numbers,sort&compress]{natbib}

\usepackage{feynmf}
\usepackage{psfrag}
\usepackage{graphicx}% Include figure files
\usepackage{bm}% bold math
\usepackage{dsfont}

\usepackage[normalem]{ulem}

\newcommand{\be}{\begin{equation}}  
\newcommand{\ee}{\end{equation}} 
\newcommand{\bal}{\begin{align}}  
\newcommand{\eal}{\end{align}}

\def\slash#1{#1\!\!\!/\!\,\,}  
\newcommand{\nl}{\nonumber \\ }
\newcommand{\order}{{\cal O}}
\newcommand{\Dslash}{D\!\!\!\!/\!\,\,}

\allowdisplaybreaks

\newcommand{\countertozero}{}
\renewcommand{\countertozero}{\setcounter{equation}{0}}
\begin{document}

\begin{titlepage}

\begin{flushright}
EFI Preprint 12-16\\
%\today
August 2, 2012
\end{flushright}

\vspace{0.7cm}
\begin{center}
\Large\bf 
Lorentz invariance in heavy particle effective theories
\end{center}

\vspace{0.8cm}
\begin{center}
{\sc Johannes Heinonen, Richard J. Hill and Mikhail P. Solon}
\\
\vspace{0.4cm}
{\it Enrico Fermi Institute and Department of Physics \\
The University of Chicago, Chicago, Illinois, 60637, USA
}

\vspace{0.4cm}
{\it E-mail:} {\tt  heinonen,richardhill,mpsolon@uchicago.edu}
\end{center}

\vspace{1.0cm}
\begin{abstract}
  \vspace{0.2cm}
  \noindent

Employing induced representations of the Lorentz group (Wigner's little group construction), 
formalism for constructing heavy particle effective Lagrangians is developed, and  
Lagrangian constraints enforcing Lorentz invariance of the $S$ matrix are derived. 
The relationship between Lorentz invariance and reparameterization 
invariance is established and it is shown why a standard ansatz for implementing 
reparameterization invariance
in heavy fermion effective Lagrangians breaks down at order $1/M^4$. 
Formalism for fields of arbitrary spin and for self-conjugate fields
is presented, and the extension to effective theories of massless fields is 
discussed. 

\end{abstract}
\vfil

\end{titlepage}

%%%%%%%%%%%%%%%%%%%%%%%%%%%%%%%%%%%%%%%%%%%%%%%%%%%%%%%%%%%%
%%%%%%%%%%%%%%%%%%%%%%%%%%%%%%%%%%%%%%%%%%%%%%%%%%%%%%%%%%%%
\section{Introduction} \label{sec:intro}
\countertozero

Heavy particle effective field theories find a wide range of applications 
in particle, nuclear and atomic physics~\cite{Caswell:1985ui,Isgur:1989vq,
Eichten:1989zv,Neubert:1993mb,Epelbaum:2008ga}.   
Recent investigations demand high orders in the $1/M$ expansion
(see e.g. \cite{Brambilla:2008zg,Hill:2011wy}), and involve construction of effective theories
for which a simple underlying ultraviolet completion is unknown, or unspecified 
(see e.g. \cite{Hill:2011be} and references therein). 
To avoid a proliferation of undetermined constants, and to enable
efficient computations, it is important to recognize that many Wilson coefficients 
are linked by Lorentz invariance to coefficients 
appearing at lower orders.  This may be viewed in analogy to the constraints 
imposed by enforcing invariance under broken chiral symmetries in low-energy chiral 
effective field theories.   
The procedure for implementing such chiral symmetry constraints, 
via the formalism of nonlinear realizations, is well known~\cite{Coleman:1969sm,Callan:1969sn}.  
It is our aim here to bring similar clarity to the implementation of Lorentz invariance in 
heavy particle effective field theories, and to provide a practical and systematic 
implementation of Lorentz
invariance constraints suitable for arbitrary orders in phenomenological applications.  

When the heavy particle is fundamental, we may derive the effective theory 
Lagrangian by introducing a field redefinition in the full theory.  For example, in terms
of an arbitrary (spacetime independent) time-like unit vector $v^\mu$, the decomposition 
of a quark field $Q(x)$ of mass $M$, 
\be
Q(x) = \label{eq:HQET}
e^{-iM v\cdot x} \left[ h_v(x) + H_v(x) \right] 
\,,
\ee
with $\slash{v}h_v = h_v$ and $\slash{v} H_v = - H_v$, 
defines an effective heavy quark field $h_v(x)$, and after integrating
out the antiparticle field $H_v(x)$, we arrive at the effective Lagrangian 
for a heavy quark.   Invariance of observables under small 
changes of $v$, so-called ``reparameterization invariance'', enforces
certain constraints on the coefficients of the effective Lagrangian~\cite{Luke:1992cs}.  
These constraints are consistent with the requirements of Lorentz invariance, 
e.g. as imposed by matching effective theory $S$ matrix elements 
to Lorentz-invariant full theory $S$ matrix elements.    
However, this construction raises several questions. 
Is reparameterization invariance a sufficient condition for
Lorentz invariance?   How do we derive a reparameterization transformation 
law without first constructing the underlying theory and 
explicitly integrating out degrees of freedom?
For applications involving a composite particle such as the proton, or 
hypothetical new particles that may not be fundamental, we cannot 
in an obvious way introduce $v$ as a parameter inside of a field 
redefinition.  What is the significance of $v$ in such cases? 
What is the general method for constructing a Lorentz invariant 
heavy particle effective field theory? 

In this paper we present the formalism of induced representations of the Lorentz group 
(Wigner's ``little group'' construction~\cite{Wigner:1939cj}) 
for application to field transformation
laws.  The parameter $v$ enters as an arbitrary reference vector in the little 
group construction.   The relationship between Lorentz invariance and reparameterization invariance
is stated precisely, and a class of allowable reparameterization transformations is obtained. 
We find that a standard ansatz for implementing reparameterization 
invariance breaks down starting at 
order $1/M^4$.
We explain this subtlety and its resolution. 

A large literature exists on topics relating to reparameterization invariance, especially
as applied to heavy quark 
Lagrangians~\cite{Kilian:1994mg,Luke:1992cs,Manohar:1997qy,Sundrum:1997ut,Finkemeier:1997re,Brambilla:2003nt,Long:2010kt,Kopp:2011gg}.   
We aim to present a conceptually clear statement of the constraints imposed
by Lorentz invariance, and of the relationship between Lorentz invariance and reparameterization 
invariance.  At a practical level, we derive explicit field transformation laws that can be 
consistently used to build Lorentz invariant Lagrangians to arbitrary order in $1/M$. 

The remainder of the paper is structured as follows.  In Section~\ref{sec:usual}
we briefly review the construction of Lorentz invariant field theories
based on finite dimensional representations of the Lorentz group. 
In Section~\ref{sec:induced} we introduce the formalism of induced 
representations and investigate the necessary conditions 
for a Lorentz invariant $S$ matrix.    
Section~\ref{sec:reparam} establishes the connection between Lorentz invariance 
and reparameterization invariance. 
A subtlety in the identification of allowable reparameterization transformations 
is explained, and a correct 
solution to the invariance equation \eqref{eq:cons2}
 is found for applications to $1/M^4$ heavy fermion 
Lagrangians. 
Section~\ref{sec:massless} provides a brief overview of 
the analogous framework for effective theories describing energetic 
massless particles.  
Section~\ref{sec:discuss} concludes with a discussion.
Appendix~\ref{sec:covariant} presents formalism in covariant notation for arbitrary spin particles 
and for self-conjugate fields. 
Appendix~\ref{sec:allorders} describes the solution of the invariance equation for 
the construction of invariant operators to arbitrary order in $1/M$. 

%%%%%%%%%%%%%%%%%%%%%%%%%%%%%%%%%%%%%%%%%%%%%%%%%%%%%%%%%%%%
%%%%%%%%%%%%%%%%%%%%%%%%%%%%%%%%%%%%%%%%%%%%%%%%%%%%%%%%%%%%
\section{Finite dimensional representations of the Lorentz algebra \label{sec:usual}}
\countertozero

The standard method for constructing Lorentz invariant Lagrangians
postulates the field transformation law
\be\label{eq:transform}
\phi_a(x) \to M(\Lambda)_{ab} \phi_b(\Lambda^{-1}x) \,,
\ee
where $M(\Lambda)$ is a finite dimensional (coordinate-independent and, in general, non-unitary) 
representation of the Lorentz group.  
In infinitesimal form, including also spacetime translations $\phi(x)\to \phi(x-a)$, we 
have 
\be\label{eq:infgen}
\delta\phi =  i ( a_0 h - \bm{a}\cdot\bm{p} - \bm{\theta}\cdot\bm{j} + \bm{\eta}\cdot\bm{k} ) \phi \,,
\ee
where $\bm{\theta}$ and $\bm{\eta}$ are infinitesimal rotation and boost parameters, 
and the generators of the Poincar\'e group acting on fields are%
\footnote{We use bold letters for euclidean three-vectors, e.g. ${\bm \partial} = ( \bm \partial^i) = (\bm \partial_i)=(\partial_x,\partial_y,\partial_z)$.} 
\begin{subequations} \label{eq:standardgen} 
\begin{align}
h &= i\partial_t \,, \label{eq:standardgenh} \\
{\bm p} &= -i {\bm \partial}  \,, \label{eq:standardgenp} \\
{\bm j} &=  {\bm r}\times {\bm p} + \bm{\Sigma} \,, \label{eq:standardgenj} \\
{\bm k} &=  {\bm r} h - t {\bm p} \pm i \bm{\Sigma} \,, \label{eq:normalkgenerator}
\end{align}
\end{subequations}
with $\Sigma^i$ the $(2s+1)$-dimensional matrix 
generators of the spin-$s$ representation of rotations
(e.g. for spin-$1/2$ Weyl fermions, 
$\bm{\Sigma}=\bm{\sigma}/2$ with $\sigma^i$ the Pauli matrices).  
Using (\ref{eq:transform}) it is straightforward to construct Lorentz invariant actions, 
and correspondingly to prove Lorentz invariance of the $S$ matrix. 
Let us briefly review this procedure.%
\footnote{For a pedagogical discussion, see \cite{Weinberg:1995mt}.}

Recall the Poincar\'e algebra
for generators of time translations $H$, space translations $P^i$, 
rotations $J^i$, and boosts $K^i$:
\begin{subequations} \label{eq:commrel}
\begin{align}
[ H, P^i ] & = 0 \,, \label{eq:CCR1}\\
[ H, J^i] &= 0 \,, \label{eq:CCR2}\\
[ P^i, P^j ] &= 0 \,, \label{eq:CCR3}\\
[ J^i, P^j ] &= i \epsilon^{ijk} P^k \,,\label{eq:CCR4} \\
[ J^i, J^j ] &= i \epsilon^{ijk} J^k \,, \label{eq:CCR5}\\
[ J^i, K^j ] &= i \epsilon^{ijk} K^k \,, \label{eq:CCR6}\\
[ H , K^i ] &= - i P^i \,. \label{eq:comm}\\
[ P^i, K^j] &= -i H \delta^{ij} \,,  \label{comm1}\\
[ K^i, K^j ] &= -i \epsilon^{ijk} J^k \,. \label{comm2} 
\end{align} 
\end{subequations}
Having built a Lagrangian that is invariant under~\eqref{eq:infgen},
we may construct the corresponding conserved charges.  Using (\ref{eq:standardgen}), 
we find in canonical quantization that these charges obey the commutation relations~(\ref{eq:commrel}).   

Lorentz invariance of the $S$ matrix demands that the free-particle charges, 
denoted by $H_0$, ${\mathbf P}_0$, ${\mathbf J}_0$, ${\mathbf K}_0$, 
commute with the scattering operator, $S=\lim_{T\to\infty} \Omega(T)^\dagger\Omega(-T)$, 
where $\Omega(T)= e^{i H T} e^{-iH_0 T}$.    
We assume that momentum and angular momentum operators for the interacting theory 
are unchanged from the free theory
and furthermore demand translational and rotational invariance of the interaction
\be\label{eq:trans}
{\bm P}={\bm P}_0 \,, \quad  
{\bm J} = {\bm J}_0\,, \quad 
[H-H_0, {\bm P}_0] = [H-H_0, {\bm J}_0] = 0 \,.
\ee  
Then
$[{\bm P}_0,S] = [{\bm J}_0,S] =0$, and by the definition of $S$ also $[H_0,S]=0$.   
Finally, if one can show~\eqref{eq:comm}
and that an asymptotic smoothness condition for $\Delta {\bm K} = {\bm K} - {\bm K}_0$ is obeyed, 
it follows that 
\begin{align}
[{\bm K}_0, S] & = \lim_{T\to\infty} [{\bm K}_0, \Omega(T)^\dagger \Omega(-T)] \\
& = \lim_{T\to\infty}\bigg\{ - [ e^{iH_0 T} \Delta {\bm K}   e^{-iH_0 T}] \Omega(T)^\dagger \Omega(-T) 
+ \Omega(T)^\dagger\Omega(-T) [ e^{-iH_0 T}  \Delta {\bm K}  e^{iH_0 T}]
\bigg\} 
=0 \, , \nonumber
\end{align}
completing the proof of the Lorentz invariance of the $S$-matrix.
For later application, we note that of the commutation relations involving $\bm{K}$, 
it is only necessary to show the relation~\eqref{eq:comm}; 
relations~\eqref{eq:CCR6}, (\ref{comm1}) and (\ref{comm2}) are not required to 
complete the proof.%
\footnote{ 
In fact, these relations 
{\it do} follow from the observation that having proven Lorentz invariance
of the $S$ matrix, it can be shown that 
$H$, ${\bm P}$, ${\bm J}$ and ${\bm K}$ are related to their free counterparts
by the  similarity transformation $\Omega(\pm \infty)$~\cite{Weinberg:1995mt}. 
}

%%%%%%%%%%%%%%%%%%%%%%%%%%%%%%%%%%%%%%%%%%%%%%%%%%%%%%%%%%%%
%%%%%%%%%%%%%%%%%%%%%%%%%%%%%%%%%%%%%%%%%%%%%%%%%%%%%%%%%%%%
\section{Effective field theory and the little group \label{sec:induced}} 
\countertozero

The field transformation law (\ref{eq:transform}), based on finite dimensional 
representations of the Lorentz group, 
is not suitable for heavy particle effective field theories.  For example, the associated  
irreducible representations of the Lorentz group are chiral, in conflict with the 
low-energy limit of a parity conserving theory such as QED or QCD. 
Let us consider instead the class of infinite dimensional induced representations.
We first review their appearance in transformations of physical states, and
then apply them as transformations acting on fields. 

%%%%%%%%%%%%%%%%%%%%%%%%%%%%%%%%%%%%%%%%%%%%%%%%%%%%%%%%%%%%
\subsection{Little group formalism \label{sec:little}} 

Consider Lorentz transformations acting on the Hilbert space of physical states for 
a spin-$s$ particle of mass $M$. 
These transformations are implemented by an 
induced representation~\cite{Wigner:1939cj}. 
In terms of  a fixed timelike 
reference vector $v^\mu$ (we assume $v^2=1$), define the associated ``little group'' as the subgroup of 
Lorentz transformations leaving $v$ invariant, $\Lambda v= v$. 
The little group for massive particles is isomorphic to $SO(3)$, the group of rotations.
Let $L(p)$ denote a standard Lorentz transformation taking $Mv$ to $p$, 
yielding a (momentum-dependent) mapping of the Lorentz group into the little group, 
\be\label{eq:WLp}
\Lambda \to W(\Lambda,p) = L(\Lambda p)^{-1} \Lambda L(p) \,. 
\ee
We may define physical states to transform schematically as 
\be \label{eq:Lorentstransform}
|p,m\rangle \to U(\Lambda,p) |p,m\rangle = \sum_{m^\prime=-s}^{s} 
D_{m^\prime\,m}[W(\Lambda,p)] |\Lambda p,m^\prime\rangle \,,
\ee
where $p^0=\sqrt{M^2+\bm{p}^2}$, and $D(W)$ is a spin-$s$ representation matrix for rotations. 
A representation for the little group thus induces a representation for 
the 
full 
Lorentz group.   

A convenient choice for the standard Lorentz transformation is $L(p) = \Lambda(p/M, v)$,	
where
$\Lambda(w,v)$ denotes the generalized rotation in the plane of the unit vectors $v$ and $w$ such that  
$\Lambda(w,v) v = w$.
This matrix is  given by
$\Lambda(w, v)  = \exp[-i\theta {\cal J}_{\alpha\beta} w^\alpha v^\beta ]$, 
with  the Lorentz generators ${\cal J}_{\alpha\beta}$  defined in Eq.~\eqref{eq:Jmunu} and
 the angle $\theta$ chosen appropriately~\cite{Luke:1992cs}.
In the vector and spinor representations
we have, respectively
\begin{subequations} \label{eq:Lambda} 
\begin{align}
\Lambda(w, v)^\mu_{\,\,\nu}  & =  
g^{\mu}_{\,\,\nu } 
- \frac{1}{ 1 + v \cdot w}\left( w^{\mu} w_\nu  + v^\mu v_\nu \right) 
+ w^{\mu} v_\nu - v^\mu w_\nu + \frac{v\cdot w}{ 1 + v\cdot w}
\left( w^{\mu} v_\nu + v^\mu w_\nu \right)\,, \\
\Lambda_{\frac12}(w,v) & = {1 + \slash{w}\slash{v} \over \sqrt{2(1+v\cdot w)} } \,. 
\end{align}
\end{subequations}
It is straightforward to verify that for elements of the little group, 
i.e. ``rotations'' with ${\cal R}v = v$,  
this choice of $L(p)$ implies
\be\label{eq:littleR}
W({\cal R},p)= {\cal R}\,,
\ee
a property that greatly simplifies the construction of invariant Lagrangians, cf.  
Sections~\ref{sec:constraints},~\ref{sec:covnot} and \ref{sec:reform} below. 
Other choices of $L(p)$ do not share this property.  
For example, suppose that we introduce a spacelike vector $s^\mu$ with $s^2=-1$.  
Then we may define $L^\prime(p)= R(p)B(p)$, with $B(p)$ a boost taking $M v^\mu$ to 
$M B(p)^\mu_{\,\,\nu} v^\nu = (v\cdot p) v^\mu + \sqrt{(v\cdot p)^2 - M^2} s^\mu$, and $R(p)$ 
a rotation taking $M B(p)^{\mu}_{\,\,\nu} v^\nu$ to $p^\mu$.  
Such an $L^\prime(p)$ provides a simple interpretation of 
 $U[L(p)]|Mv,m\rangle$ in terms of helicity eigenstates (note that the spacelike vector is required
to define a direction for helicity decomposition), but this consideration is
secondary to the simplicity of (\ref{eq:littleR}) for our present purposes. 

The remaining independent Lorentz generators represent ``boosts'' that
shift $v$.   
They can be chosen as ${\cal B}(q) = \Lambda(v-q/M, v)$ with $(v-q/M)^2=1$.
The appearance of the $1/M$ factor in $v-q/M$ will be explained in Section~\ref{sec:constraints} below.
For an infinitesimal momentum $q$, which obeys $v\cdot q= \order(q^2)$, these boosts are given by
\begin{subequations} \label{eq:Bboost} 
\begin{align}
{\mathcal B}(q)^\mu_{\,\,\nu}  & =  
g^{\mu}_{\,\,\nu } 
+ \frac{v^{\mu} q_\nu  - q^\mu v_\nu}{M} + {\mathcal O}(q^2)  \,, \\
{\mathcal B}_{\frac12}(q) & = {1 - \frac{\slash{q}\slash{v}}{2M}  } + {\mathcal O}(q^2) \,. 
\end{align}
\end{subequations}
For the transformation~\eqref{eq:Lorentstransform}, we find 
\begin{equation}\begin{split} \label{eq:WB}
W({\cal B}(q),p) 
& = 1 - {i\over 2} \left[ \frac{1}{M(M+v\cdot p)} ( q^\alpha p_{\perp}^{\beta} - p_\perp^\alpha q^\beta ) \right] {\cal J}_{\alpha\beta}
+ \order(q^2) ,
\end{split}\end{equation}
where for any four-vector $k$ we define $k_\perp^\mu \equiv k^\mu - (v\cdot k) v^\mu$. 

%%%%%%%%%%%%%%%%%%%%%%%%%%%%%%%%%%%%%%%%%%%%%%%%%%%%%%%%%%%%
\subsection{Field transformation law and Lorentz invariance}

In place of (\ref{eq:transform}) let us postulate the transformation
law for free massive fields,
\begin{equation} \label{eq:little_transform}
\phi_a(x) \to D[ W(\Lambda, i\partial ) ]_{ab} \phi_b(\Lambda^{-1}x) \,. 
\end{equation}
For notational simplicity consider  the special
choice $v=(1,0,0,0)$.  
Equation~\eqref{eq:little_transform} together with Eq.~\eqref{eq:WB} corresponds 
to replacing the boost generator~\eqref{eq:normalkgenerator} by%
\footnote{
For spin-$1/2$ particles, (\ref{eq:infin}) may also be obtained by 
performing a Foldy-Wouthuysen 
transformation on Eq.~\eqref{eq:normalkgenerator}~\cite{Foldy:1956self}.}
\begin{align}\label{eq:infin}
\bm{k} &= {\bm r} h  -  t \bm{p}  
\pm i { \bm{\Sigma} \times \, \bm{\partial} \over M + \sqrt{M^2-\bm{\partial}^2}} \,. 
\end{align}
The generators~\eqref{eq:standardgenh}-\eqref{eq:standardgenj} 
together with~\eqref{eq:infin} will satisfy the Poincar\'e algebra when acting on fields satisfying 
\be
i\partial_t \phi = \pm \sqrt{M^2-\bm{\partial}^2} \phi \,. 
\ee
It follows that the conserved charges derived from a free field 
Lagrangian invariant under (\ref{eq:little_transform}) will satisfy (\ref{eq:commrel}). 

In contrast to (\ref{eq:transform}), 
transformation (\ref{eq:little_transform}) acts on the 
field coordinates, spoiling gauge invariance. 
To include gauge interactions, we promote the partial derivatives in (\ref{eq:little_transform}) to 
covariant derivatives $D_\mu = \partial_\mu-ig A_\mu^A t^A \equiv \partial_\mu - i g A_\mu$,
\be\label{eq:cov}
\phi_a(x) \to D[ W(\Lambda, iD) ]_{ab} \phi_b(\Lambda^{-1}x) \,,
\ee
and correspondingly the infinitesimal generators become 
\begin{subequations} \label{eq:infincov}
\begin{align}
h &= i\partial_t \,, \label{eq:infinh}\\
{\bm p} &= -i {\bm \partial}  \,, \label{eq:infinp}\\
{\bm j} &= {\bm r}\times \bm{p} + \bm{\Sigma} \,, \label{eq:infinj}\\
{\bm k} &= {\bm r} h - t {\bm p} 
\pm i { \bm{\Sigma} \times {\bm D} \over M 
+ \sqrt{M^2-\bm{D}^2}} + \order(g) \,. \label{eq:kwithcovd}
\end{align}
\end{subequations} 
In the expansion of  ${\bm D}/(M + \sqrt{M^2-\bm{D}^2})$
we assume a choice of ordering for the covariant derivatives.
The $\order(g)$ terms in ${\bm k}$ denote field strength-dependent corrections 
that vanish for the non-interacting theory (i.e. $g\to 0$).    
Such $\order(g)$ terms can be introduced so that the resulting invariant Lagrangian is 
in ``canonical form'', i.e. where the only time derivative acting on $\phi$ 
appears in the leading term,
\be\label{eq:canonical}
{\cal L} = \bar{ \phi} ( iD_t + \dots )\phi \,. 
\ee
The existence of suitable field strength-dependent terms, ensuring
a boost generator $\bm{k}$
which yields a non-zero invariant Lagrangian,
is implied by the all-orders construction 
in Section~\ref{sec:reparam} and Appendix~\ref{sec:allorders}.
The explicit form of these corrections is not required for the following argument.  

Although the field-dependent generators (\ref{eq:infincov}) do not obey simple 
commutation relations, 
we may nevertheless show that the $S$ matrix derived from the 
resulting invariant action
is Lorentz invariant (and hence that the conserved charges in the interacting theory 
satisfy the Poincare algebra). 
To see this, we assume as before the relations (\ref{eq:trans}).    
Relation (\ref{eq:comm}) is satisfied if the explicit time dependence of 
the conserved charge $\bm{K}$ satisfies $\partial\bm{K}/\partial t= -\bm{P}$, so that 
\be\label{eq:Kconserve}
0 = {d\over dt} \bm{K} = {\partial\over\partial t} \bm{K} + i[H, \bm{K}] = - \bm{P} + i [H, \bm{K}] \,.  
\ee
The  fact that $\partial\bm{K}/\partial t= -\bm{P}$ follows from 
the assumed form of the infinitesimal generators (\ref{eq:infincov}).  
For the boost 
$\phi \to (1 + i \bm{\eta} \cdot \bm{k}) \phi$, we find the  conserved charge%
\footnote{
The first ellipsis in (\ref{eq:Kt}) includes possible contributions from 
a surface term in $\delta {\cal L}$, which do not affect the term with 
explicit $t$ dependence in (\ref{eq:Kt}). 
}
\begin{align} \label{eq:Kt}
\bm{K}
& = \sum_\phi i  \int d^3x  \frac{\delta \mathcal{L}}{\delta \dot \phi} \,  \bm{k}\,  \phi  + \ldots 
=  \sum_\phi i  \int
d^3x  \frac{\delta \mathcal{L}}{\delta \dot \phi} \, 
\left[ - t \bm{p} \right]  \phi + \ldots 
= -t \bm{P} + \ldots \,.
\end{align} 
Here the important point is that 
the remaining terms have no \emph{explicit} time dependence, so 
that (\ref{eq:Kconserve}) follows. 

Let us close this section with two comments. First,  
the choice $v=(1,0,0,0)$ is not essential to the argument. 
The generators for arbitrary $v$ can be obtained by a coordinate 
change using a boost which takes $(1,0,0,0)$ to $v$.
While the resulting explicit expressions for rotation and boost generators 
become more complicated, the demonstration of Lorentz invariance is not essentially
changed. 
Second, having specified an ordering for covariant derivatives   appearing in 
the boost generator $\bm{k}$, additional field strength-dependent corrections are determined at each order in $1/M$ by
enforcing that the resulting invariant Lagrangian is in canonical form.   We illustrate this with
an explicit example in the following subsection.  The existence of such a generator is implied 
by the analysis of Section~\ref{sec:reparam} and Appendix~\ref{sec:allorders}.

%%%%%%%%%%%%%%%%%%%%%%%%%%%%%%%%%%%%%%%%%%%%%%%%%%%%%%%%%%%%
\subsection{$1/M$ expansion and Lagrangian constraints \label{sec:constraints}}

To enable the $1/M$ expansion we extract the rest mass by the field redefinition, 
\be\label{eq:phase}
\phi(x) = e^{-iMt} \phi^\prime(x) \,.
\ee
In phenomenological applications it 
 is also convenient to work with non-relativistic field normalization 
\be\label{eq:norm}
\phi^\prime(x) = 
\left( M^2 \over M^2-\bm{D}^2 \right)^\frac14 \phi^{\prime\prime}(x) \,. 
\ee 
We enforce invariance under (\ref{eq:infinh}), (\ref{eq:infinp}) and (\ref{eq:infinj}) 
by ensuring translational invariance (no explicit factors 
of $x^\mu$) and rotational invariance.
For the boost transformation (\ref{eq:kwithcovd}) 
we use $\bm{\eta}= -\bm{q}/M$ in (\ref{eq:infgen}) to 
preserve the power counting $D_t = \order(1/M)$ in (\ref{eq:deriv}). 
This explains the appearance of $1/M$ in (\ref{eq:Bboost}). 
The resulting $1/M$ expansion becomes%
\footnote{For notational clarity we leave the coordinate 
change $x \to x'= {\cal B}^{-1} x$ implicit and suppress 
primes on coordinates and derivatives in (\ref{eq:lorentz}) and (\ref{eq:deriv}).}
\begin{align}\label{eq:lorentz}
\phi^{\prime\prime} 
&\to e^{-i\bm{q}\cdot\bm{x}}\left\{  
1 + {i\bm{q}\cdot\bm{D}\over 2 M^2} + {i \bm{q}\cdot\bm{D} \bm{D}^2 \over 4 M^4} 
- 
{ \bm{\Sigma}\times \bm{q} \cdot \bm{D} \over 2 M^2} 
\left[ 1 + {\bm{D}^2\over 4 M^2} \right] + \order(g,1/M^5) 
\right\} \phi^{\prime\prime}\, .
\end{align}
Gauge fields are assumed to transform as usual, in the vector representation 
of the Lorentz group.   Combined with derivatives acting on the transformed 
coordinate in (\ref{eq:lorentz}), we have 
\be\label{eq:deriv}
D_t \to D_t + {1\over M} \bm{q}\cdot\bm{D} \,, \quad
\bm{D} \to \bm{D} + {1\over M} \bm{q} D_t  \,.  
\ee

To illustrate the constraints, consider the canonical form of the 
abelian gauged heavy spin-$1/2$ fermion effective Lagrangian (i.e., NRQED)
through $\order(1/M^3)$. Identifying $\phi^{\prime\prime} = \psi$ as a two-component spinor 
and setting $g=-e$ we obtain~\cite{Kinoshita:1995mt,Manohar:1997qy} 
\begin{align}\label{eq:abelian}
  {\cal L} &= \psi^\dagger
  \bigg\{  i D_t  + c_2 {\bm{D}^2 \over 2M}  + c_4 {\bm{D}^4 \over 8 M^3} +
  c_F e{ \bm{\sigma}\cdot \bm{B} \over 2M}   
+ c_D e{ [\bm{\partial}\cdot \bm{E}] \over 8 M^2}  
+ i c_S e{ \bm{\sigma}
    \cdot ( \bm{D} \times \bm{E} - \bm{E}\times \bm{D} ) \over 8 M^2} 
\nl
&\quad
+ c_{W1}e {  \{ \bm{D}^2 ,  \bm{\sigma}\cdot \bm{B} \}  \over 8 M^3}  
- c_{W2}e {  D^i \bm{\sigma}\cdot
    \bm{B} D^i \over 4 M^3 }  
    + c_{p^\prime p} e { \bm{\sigma} \cdot
    \bm{D} \bm{B}\cdot \bm{D} + \bm{D}\cdot\bm{B} \bm{\sigma}\cdot \bm{D}
    \over  8 M^3} 
\nl
&\quad 
+ i c_M e { \{ \bm{D}^i ,  [\bm{\partial} \times \bm{B}]^i \} \over 8 M^3}
+ c_{A1} e^2{ \bm{B}^2 - \bm{E}^2 \over 8 M^3} 
 - c_{A2} e^2{ \bm{E}^2 \over 16 M^3 } 
+ \order(1/M^4) \bigg\} \psi \,. 
\end{align}
Here we have defined  $E^i = (-i/e)[D_t, D^i]$, $\epsilon^{ijk} B^k \equiv (i/e)[D^i, D^j]$.
Under (\ref{eq:lorentz}), a straightforward computation yields 
\be \label{eq:LIexplicit}
\delta {\cal L} = {1\over M} \delta {\cal L}_1 + {1\over M^2} \delta {\cal L}_2 
+  {1\over M^3} \delta {\cal L}_3
+ \dots \,,
\ee
where using $\bm{\Sigma}=\bm{\sigma}/2$ in (\ref{eq:lorentz}),
\begin{subequations}
\begin{align} 
\delta {\cal L}_1 &= \psi^\dagger \left[  (1- c_2) i\bm{q}\cdot \bm{D}  \right] \psi \,, \\
\delta {\cal L}_2 &= \psi^\dagger \left[ 
-\frac12(1-c_2) \{ \bm{q}\cdot\bm{D} , D_t \} 
+ \frac{e}{4} ( 1 - 2 c_F + c_S ) \bm{\sigma}\times \bm{q} \cdot \bm{E} 
   \right] \psi \,, \\
\delta {\cal L}_3 &= \psi^\dagger \bigg[ 
 {e \over 8} c_D [ D_t , \bm{q}\cdot\bm{E} ] 
+ \frac{e}{8} \left( c_F - c_D + 2 c_M \right)  \bm{q}\cdot [\bm{\partial}\times \bm{B}] 
+ {i\over 4} \left( c_2 - c_4 \right)  \{ \bm{q}\cdot\bm{D}, \bm{D}^2 \}
\\
&\quad
+ {i e \over 8} c_S \{ D_t, \bm{\sigma}\times\bm{q}\cdot \bm{E} \}
+ {ie \over 8} \left( c_2 +2c_F - c_S -2c_{W1} + 2c_{W2}  \right) \{\bm{q}\cdot \bm{D} , \bm{\sigma}\cdot \bm{B} \} 
\nl
&\quad
+ {i e \over 8}\left( -c_2 + c_F - c_{p^\prime p} \right) \{\bm{\sigma}\cdot \bm{D} , \bm{q}\cdot \bm{B} \} 
+ {i e \over 8}  \left(-c_F + c_S - c_{p^\prime p} \right)  \bm{q}\cdot\bm{\sigma}(\bm{D}\cdot\bm{B}+\bm{B}\cdot\bm{D}) 
\nonumber
\bigg] \psi \,.
\end{align} 
\end{subequations}
From $\delta {\cal L}_1$ and $\delta {\cal L}_2$, we find
\begin{align}\label{eq:constrain12}
c_2 &= 1 \,, \quad
 c_S = 2 c_F - 1 \,.
\end{align}
The variation $\delta {\cal L}_3$ is equivalent to zero 
upon a field strength-dependent modification of the boost transformation (\ref{eq:lorentz}),  
\be\label{eq:redef}
\psi(x) \!\to\! 
 e^{-i\bm{q}\cdot\bm{x}}\!\left\{\!
1 + {i\bm{q}\cdot\bm{D}\over 2 M^2} 
- 
{ \bm{\sigma}\times \bm{q} \cdot \bm{D} \over 4 M^2} 
+ {i c_D \over 8 M^3} e \bm{q}\cdot \bm{E} 
+ {c_S \over 8 M^3} e \bm{q}\cdot\bm{\sigma}\times\bm{E} 
+ \order\left(1\over M^4\right) 
\right\} 
\psi({\cal B}^{-1} x)  \,,
\ee
and upon enforcing the constraints~\cite{Hill:2011wy,Manohar:1997qy}
\begin{align}\label{eq:constrain3}
 c_4 &= 1 \,, \quad
2c_M = c_D - c_F \,, \quad
c_{W2} = c_{W1} - 1 \,, \quad 
c_{p^\prime p} = c_F - 1 \,.  
\end{align}
The computation of the complete Lagrangian at $\order(1/M^4)$ is presented in \cite{nrqedm4}. 

%%%%%%%%%%%%%%%%%%%%%%%%%%%%%%%%%%%%%%%%%%%%%%%%%%%%%%%%%%%%
%%%%%%%%%%%%%%%%%%%%%%%%%%%%%%%%%%%%%%%%%%%%%%%%%%%%%%%%%%%%
\section{Reparameterization invariance and invariant operators\label{sec:reparam}}
\countertozero

While in practice it may be convenient to enforce Lorentz invariance only
after expanding the Lagrangian in a series of rotationally-invariant, but not
Lorentz invariant, operators, it is interesting to consider formalism that
permits an explicitly Lorentz-invariant construction.    
This formalism also addresses the question of existence of a suitable boost
generator, extending (\ref{eq:redef}) to arbitrary order in $1/M$. 

This section begins by introducing covariant notation that can either be used in place of the
$v=(1,0,0,0)$ formalism above, or used to construct manifestly invariant operators.  
The relation between Lorentz invariance and reparameterization invariance is then demonstrated, 
and a general discussion of the invariant operator method is presented. 
In particular, we derive the necessary invariance equation~\eqref{eq:cons2} and 
present the solution to order $1/M^3$.
A systematic, all-orders solution of the invariance equation is given in Appendix~\ref{sec:allorders}. 

%%%%%%%%%%%%%%%%%%%%%%%%%%%%%%%%%%%%%%%%%%%%%%%%%%%%%%%%%%%%
\subsection{Covariant notation \label{sec:covnot}} 

The formalism of Appendix~\ref{sec:covariant} allows us to straightforwardly 
extend the discussion to a general reference vector $v$ and to 
arbitrary spin.
Consider a term in the Lagrangian of the schematic form 
\be\label{eq:schematic}
\bar{\phi}_v \bigg\{ \cdots v^\mu  \cdots D^\mu \cdots \gamma^\mu \cdots \bigg\} \phi_v \,,
\ee 
where indices are contracted with $g_{\mu\nu}$ and $\epsilon_{\mu\nu\rho\sigma}$.
Invariance under generalized rotations of such a term in the action 
follows using the field 
transformation (\ref{eq:littleR}),
\be\label{eq:rot1}
\phi_v(x) \to {\cal R} \phi_v(x^\prime) \,,
\ee
where
$x^\prime \equiv {\cal R}^{-1}x$. 
The transformation of the derivative and the gauge field are as usual,
\begin{align}\label{eq:rot2}
\partial^\mu \to \partial^{\mu} = \mathcal{R}^{\mu}_{\,\,\nu} \partial^{\prime\nu} \,, 
\quad 
A^{ \mu} \to \mathcal{R} ^{\mu}_{\,\,\nu} A^{ \nu}(x^\prime) \,.
\end{align}
If the Lagrangian is already constructed such that all vector and spinor indices are contracted in~\eqref{eq:schematic}, we can easily see that the Lagrangian is invariant under generalized rotations using the identities 
\be\label{eq:rot3}
v^\mu = {\cal R}^\mu_{\,\,\nu}v^\nu 
\,, \qquad 
\gamma^\mu = {\cal R}_{\frac12} \left(  {\cal R}^{\mu}_{\,\,\nu} \gamma^\nu \right) {\cal R}_{\frac12}^{-1} 
\,.
\ee 
According to~\eqref{eq:WB}, the infinitesimal boosts are implemented by
\be\label{eq:covboost}
\phi_v(x) \to W({\cal B}, iD) \phi_v( x^\prime) 
\,, 
\ee 
where $ x^\prime \equiv {\cal B}^{-1}x$, 
together with the transformation of the derivative and gauge field, 
\begin{align}\label{eq:covderiv}
\partial^\mu \to \partial^{\mu} = \mathcal{B}^{\mu}_{\,\,\nu}  \partial^{\prime\nu} \,, 
\quad 
A^{ \mu}(x) \to \mathcal{B} ^{\mu}_{\,\,\nu} A^{ \nu}( x^\prime) \,.
\end{align}
We may proceed as in Section~\ref{sec:constraints}
above to construct invariant combinations of Lagrangian interactions of the form 
(\ref{eq:schematic}), order by order in $1/M$. 

As an explicit example, let us focus presently on the phenomenologically important 
one-heavy particle sector of a spin-$1/2$ theory.   
To enable the $1/M$ expansion and convert to non-relativistic normalization, 
we introduce the field redefinition as in~\eqref{eq:phase} and~\eqref{eq:norm},
\be \label{eq:phasenN}
\psi_v(x) = e^{-iMv\cdot x}
N(v,iD)
\psi^\prime_v(x) \,,
\quad
N(v,iD) = \left( {M^2 \over M^2 + D_\perp^2 } \right)^{\frac14} \,.
\ee
The boost transformation (\ref{eq:covboost}) becomes 
\be
\psi_v^\prime \to e^{iq\cdot x} 
\tilde{W}_\frac12({\cal B}, iD+ M v) 
\psi_v^\prime 
\,, 
\ee
where 
\be
\tilde{W}({\cal B}, iD+ M v)
=
N(v+q/M,iD-q)^{-1} 
W({\cal B}, iD+ M v) N(v,iD)  \,.
\ee
The $1/M$ expansion of this transformation is the extension to arbitrary $v$, for spin-$1/2$, 
of the previous (\ref{eq:lorentz}):
\be \label{eq:covlorentz}
\psi_v^\prime \to 
e^{iq\cdot x} \bigg\{ 
1 + {iq\cdot D_\perp \over 2 M^2} - {i q\cdot D_\perp D_\perp^{2} \over 4 M^4} 
+ {1\over 4 M^2} \sigma_{\alpha\beta} q^\alpha D_\perp^{\beta} 
\left[ 1 - {D_\perp^{2}\over 4 M^2} \right] + \order(g,1/M^5) \bigg\} \psi_v^\prime \,.
\ee
Similarly, we find the extension to arbitrary $v$ of the transformations \eqref{eq:deriv}
\begin{equation}
v \cdot D \to v \cdot D + \frac{1}{M} q \cdot D_\perp 
\,, \quad
D_\perp^\mu \to D_\perp^{\mu} - \frac{1}{M} q^\mu (v \cdot D).
\end{equation}
Using these transformations one can build an invariant Lagrangian, which 
(in the abelian case) is equivalent to the extension of the 
Lagrangian~\eqref{eq:abelian} to arbitrary $v$ with the same 
constraints~\eqref{eq:constrain12} and~\eqref{eq:constrain3}.

%%%%%%%%%%%%%%%%%%%%%%%%%%%%%%%%%%%%%%%%%%%%%%%%%%%%%%%%%%%%
\subsection{Reparameterization invariance \label{sec:reform}} 

We can reformulate  the transformation law for generalized boosts 
by using the identities, 
\be
v^\mu = {\cal B}^\mu_{\,\,\nu} ({\cal B}^{-1})^{\nu}_\rho v^\rho \equiv {\cal B}^{\mu}_{\nu} w^\nu  \,, \quad
\gamma^\mu = {\cal B}_{\frac12} \left(  {\cal B}^{\mu}_{\,\,\nu} \gamma^\nu \right) {\cal B}_{\frac12}^{-1} \,.
\ee 
In place of (\ref{eq:covboost}) and (\ref{eq:covderiv}) the 
transformation of any operator of the form (\ref{eq:schematic}) is 
identical to the transformation obtained by the substitutions
\be\label{eq:covrep}
v \to w = v + q/M \,, 
\quad  
\phi_v \to \phi_w \equiv 
{\cal B}^{-1} W({\cal B}, iD_\mu) \phi_v \, ,
\ee
with no transformation of the coordinate and gauge field. 
The rules (\ref{eq:covrep}), with suitable choice for $W$, may be identified with 
the rules obtained by enforcing ``reparameterization invariance''~\cite{Luke:1992cs}. 
However, we emphasize that from the present perspective, we are not changing the reference 
vector $v$, but simply noticing the equivalence of 
(\ref{eq:covboost}) and (\ref{eq:covderiv}) on the one hand,
and (\ref{eq:covrep}) on the other hand, when acting on operators of the form (\ref{eq:schematic}). 

%%%%%%%%%%%%%%%%%%%%%%%%%%%%%%%%%%%%%%%%%%%%%%%%%%%%%%%%%%%%
\subsection{Invariant operator method}

It is not obvious that a non-zero Lagrangian, invariant
under~\eqref{eq:covboost} and~\eqref{eq:covderiv} to arbitrary order,  will exist.  
For example, in (\ref{eq:LIexplicit}) invariance relies on
the possibility to enforce $\delta {\cal L}_n=0$ by modifying
the boost generator as in (\ref{eq:redef}) and 
enforcing relations as in \eqref{eq:constrain12}
and \eqref{eq:constrain3}.  
It is not evident that this procedure can be extended to arbitrary order. 
We present here a method of constructing 
operators that are manifestly invariant under a particular choice of boost 
generator, to arbitrary order in $1/M$.
The details of the construction are given in
Appendix~\ref{sec:allorders}.

The embedding of the little group into constrained representations of the 
full Lorentz group (cf. Appendix~\ref{sec:covariant}) provides a framework 
for constructing explicitly invariant operators. 
Suppose that we find an operator $\Gamma(v,iD)$
such that 
\be \label{eq:cons}
\Gamma(\Lambda^{-1} v, iD) \Lambda^{-1} W(\Lambda, iD) = \Gamma(v, i D) \,.
\ee
when acting on fields $\phi_v$ obeying the appropriate constraints, as given in Appendix~\ref{sec:covariant} 
(e.g. $\slash{v}\phi_v = \phi_v$ for spin-$1/2$). 
It follows from the rules (\ref{eq:covrep}) that the combination
\be
\Phi_v \equiv \Gamma(v,iD) \phi_v
\ee
is invariant under the reparameterization implementation (\ref{eq:covrep}) of 
generalized boosts.    Provided that invariance under generalized
rotations (\ref{eq:rot1})-(\ref{eq:rot3}) 
is maintained, we may build operators that are explicitly invariant. 
For example, in the spin-$1/2$ case
\be
\bar{\Psi}_v i\Dslash \Psi_v \,, \quad \bar{\Psi}_v \Psi_v \,, \quad
\bar{\Psi}_v i\sigma^{\mu\nu} [D_\mu,D_\nu] \Psi_v \,,  
\ee
are invariant. 
Note that because of Eq.~\eqref{eq:littleR} the only constraints 
on $\Gamma(v,iD)$  from Eq.~\eqref{eq:cons} come from boosts $\Lambda=\cal B$.

Applying field redefinitions as in \eqref{eq:phasenN}, 
the condition (\ref{eq:cons}) for $\Gamma$ becomes
\be \label{eq:cons2}
\Gamma(v+q/M, iD-q)  
{\cal B}^{-1} 
 \tilde{W}({\cal B}, iD + M v)
 = \Gamma(v, i D) \,.
\ee
We will refer to (\ref{eq:cons2}) as the ``invariance equation''.
Provided that such a $\Gamma(v,iD)$ can be found, the field
\be
\Phi_v^\prime(x) \equiv \Gamma(v,iD) \phi^\prime_v(x) 
\ee
obeys a simple transformation law under the reparameterization implementation 
of generalized boosts (\ref{eq:covrep}),
\be
\Phi_v^\prime \to \Phi_w^\prime \equiv e^{iq\cdot x} \Phi_v^\prime \,.
\ee 
Noting that  
$e^{-iq\cdot x} 
( iD^\mu+ M w^\mu ) e^{iq\cdot x} 
=
iD^\mu+ M v^\mu$,
invariant operators may thus be built from contractions of 
polynomials of $\gamma^\mu$ and $v^\mu + i D^\mu/M$, 
between $\bar{\Phi}_v^\prime$ and $\Phi_v^\prime$.  
For example in the spin-$1/2$ case, 
\be
\bar{\Psi}_v^\prime ( i\Dslash + M\slash{v} ) \Psi_v^\prime \,, \quad 
\bar{\Psi}_v^\prime \Psi_v^\prime \,, \quad
\bar{\Psi}_v^\prime i \sigma^{\mu\nu} [D_\mu, D_\nu] \Psi_v^\prime \,, 
\ee
are invariant. 

%%%%%%%%%%%%%%%%%%%%%%%%%%%%%%%%%%%%%%%%%%%%%%%%%%%%%%%%%%%%
\subsection{Solution for $\Gamma(v,iD)$ \label{sec:gammasol}} 

The key element of the invariant operator construction 
is a solution of the invariance equation (\ref{eq:cons2}). 
Without loss of generality, let us set $N(v,iD)=1$;
the  solution for general $N$ can then be obtained by 
$\Gamma(v,iD) \to  \Gamma(v,iD) N(v,iD)^{-1}$. 
The method presented can be easily extended to arbitrary spin, but 
for illustration we focus on the one-heavy particle sector of a spin-$1/2$ theory. 

In order to obtain a solution in closed form for the free theory, and to 
make contact with previous work, 
it is convenient to 
take the free theory limit for $W_{\frac12}({\cal B},i\partial+Mv)$ of the form~\cite{Luke:1992cs}%
\begin{align}\label{eq:Wfree}
W_{\frac12}({\cal B},i\partial+Mv)
&= 
{\cal B}_{\frac12} \,\Lambda_{\frac12}(\hat{\cal V}_{\rm free},v+q/M)^{-1} \Lambda_{\frac12}(\hat{\cal V}_{\rm free},v)
\\
&= 
1 + 
{1\over 4 M^2}\sigma_\perp^{\mu\nu}q_\mu \partial_\nu \left[ 
 1 - {iv\cdot \partial\over M} + {1\over M^2}\left( (iv\cdot\partial)^2 -\frac14 (i\partial_\perp)^2 
\right)
\right]
+\order(1/M^5)
\,, \nonumber
\end{align}
where $\Lambda_{\frac12}(u,v)$ was defined in (\ref{eq:Lambda}), 
${\cal V}_{\rm free}^\mu \equiv v^\mu + i \partial^\mu/M$ and 
$\hat{\cal V}_{\rm free}^\mu \equiv {\cal V}_{\rm free}^\mu/|{\cal V}_{\rm free}|$. 
We have also used that $\slash v \psi_v = \psi_v$. 
Inspection of (\ref{eq:cons2}) shows that 
an all-orders solution can be written for $\Gamma$ in the non-interacting theory, 
\begin{equation} \begin{split}\label{eq:gammafree}
\Gamma(v,i\partial) &= \Lambda_{\frac12}(\hat{\cal V}_\text{free}, v) 
= 1 + {i\slash{\partial}_\perp\over 2 M} + {1\over M^2}\left[ -\frac18 (i\partial_\perp)^2 
- \frac12 i \slash{\partial}_\perp iv\cdot\partial 
\right] \\
&\quad
+ {1\over M^3} \left[ \frac14 (i\partial_\perp)^2 iv\cdot\partial 
+  {i\slash{\partial}_\perp \over 2}\left( 
- {3\over 8} (i\partial_\perp)^2  
+ (iv\cdot \partial)^2 
\right) 
\right] + \order(1/M^4) 
\,.
\end{split}\end{equation}
In the interacting theory it turns out that one cannot 
simply replace $\partial$ by $D$ in~\eqref{eq:gammafree}
to obtain a solution
for $\Gamma(v,iD)$.  
It is instead necessary to add specific field strength dependent terms, 
first to $W$ (as in (\ref{eq:Wchoice}) and (\ref{eq:defofX}) below) 
in order to satisfy consistency conditions, 
and then to $\Gamma$ in order to solve the invariance
equation~\eqref{eq:cons2}.  The computations of
Appendix~\ref{sec:allorders} show that a solution for $\Gamma(v,iD)$ will exist if 
we specify
\begin{align} \label{eq:Wchoice}
&W_{\frac12}({\cal B}, iD+Mv) = 
1 + {1\over 4M^2} \sigma^\perp_{\mu\nu} q^\mu D_\perp^\nu 
\left(1 - {iv\cdot D \over M} \right) + \order(1/M^4)\,,
 \end{align}
with (\ref{eq:Wchoice}) reducing to (\ref{eq:Wfree}) at $g=0$.
Let us proceed through $\order(1/M^3)$, writing 
\be\label{eq:gammaexpand}
\Gamma = 1 + {1\over M} \Gamma^{(1)} + {1\over M^2}\Gamma^{(2)} + {1\over M^3} \Gamma^{(3)} 
+ \dots \,,
\ee
and deriving a solution to the invariance equation~\eqref{eq:cons2} order by order in $1/M$.
In Appendix~\ref{sec:allorders} 
we present a systematic construction that extends the solution to arbitrary order.

Modulo terms that vanish when acting on $\psi_v$ with $\slash{v}\psi_v=\psi_v$, 
we find 
\begin{subequations} \label{eq:gammaint}
 \begin{align}
  \quad \Gamma^{(1)} & = \frac12 i\Dslash_\perp  \,.  \label{eq:gamma01} \\
  \Gamma^{(2)} & = -\frac18 (iD_\perp)^2 -\frac12 i \Dslash_\perp iv\cdot D 
 + g A \sigma^{\mu\nu} G_{\mu\nu} + g B \gamma^\mu v^\nu G_{\mu\nu}  \,. \label{eq:gamma2}\\
 \Gamma^{(3)} &  = \frac14 (iD_\perp)^2 iv\cdot D  + {i\Dslash_\perp \over 2} \left[
-\frac38 (i D_\perp)^2 + (iv\cdot D)^2 \right] 
-{g\over 8} G_{\mu\nu} v^\mu D_\perp^\nu 
-\frac{g}{16} \sigma_\perp^{\mu\nu} G_{\mu\nu} i\Dslash_\perp
 \,, \label{eq:sol}
 \end{align}
 \end{subequations}
where we define $[iD_\mu , iD_\nu] = ig G_{\mu\nu}$. 
Starting at order $1/M^2$ the solution is not unique. 
However, since we will consider arbitrary factors of 
${\cal V}^\mu \equiv  v^\mu + iD^\mu/M$ when constructing invariant operators, 
we can set $A=B=0$ by considering instead of $\Gamma$, the operator $\Gamma^\prime$ given by 
\be\label{eq:field}
\Gamma(v,iD) = \left( 1 - {iA} \sigma_{\mu\nu} [{\cal V}^\mu , {\cal V}^\nu ] 
- {iB} \gamma_\mu {\cal V}_\nu [{\cal V}^\mu , {\cal V}^\nu ]  + \dots \right) \Gamma^\prime(v,iD) \,. 
\ee
Similarly, we have absorbed additional $1/M^3$ terms in~\eqref{eq:sol}. 
The remaining terms in (\ref{eq:gammaint}) have free derivatives $D_\mu$ acting to the right, 
and cannot be removed as in (\ref{eq:field}). 

A complete basis of bilinears required through order $1/M^3$ is
\begin{align}\label{eq:Lfull}
{\cal L} &= \bar{\Psi}_v \bigg\{ 
M(\slash{\mathcal{V}}-1) 
- a_F g {\sigma^{\mu\nu} G_{\mu\nu} \over 4 M}  
+ i a_D g { \{ {\cal V}_\mu , [M{\cal V}_\nu, G^{\mu\nu} ] \} \over 16 M^2} 
- a_{W1} g { [M{\cal V}^\alpha, [M{\cal V}_\alpha, \sigma^{\mu\nu} G_{\mu\nu} ] ] \over 16 M^3} 
\nl
&\qquad\qquad
+ a_{A1} g^2 { G_{\mu\nu}G^{\mu\nu} \over 16 M^3} 
+ a_{A2} g^2 { {\cal V}_\alpha G^{\mu\alpha} G_{\mu\beta} {\cal V}^\beta \over 16 M^3}  
\bigg\}\Psi_v  \,. 
\end{align}
Performing field redefinitions to arrive at canonical form, we recover the 
result (\ref{eq:abelian}) with constraints (\ref{eq:constrain12}) and 
(\ref{eq:constrain3}).   
The computation at $\order(1/M^4)$ is presented in \cite{nrqedm4}. 
We may perform a similar computation for heavy vector particles (or 
particles of arbitrary spin), and/or enforce constraints appropriate to self-conjugate
fields (cf. Appendix~\ref{sec:covariant}). 

The passage from (\ref{eq:gammafree}) to%~(\ref{eq:gamma01}),(\ref{eq:gamma2}),(\ref{eq:sol}) %
~\eqref{eq:gammaint} is not as simple as previously envisaged~\cite{Luke:1992cs,Manohar:1997qy},
and careful attention must be paid to the interplay of Lorentz and gauge symmetry. 
The computations in Appendix~\ref{sec:allorders} show that an arbitrary 
``covariantization'' of (\ref{eq:Wfree}) does {\it not} solve the invariance 
equation (\ref{eq:cons2}).
The covariant little group element $W({\cal B},iD+Mv)$ must satisfy consistency 
conditions for a solution to exist, and
specific field strength dependent terms, such as those appearing in~\eqref{eq:sol}, 
are necessary in order that $\Gamma(v,iD)$ 
satisfy the resulting invariance equation~\eqref{eq:cons2}.
These considerations have previously been overlooked~\cite{Luke:1992cs,Manohar:1997qy}. 
For example, a naive covariantization of Eq.~\eqref{eq:gammafree}, 
\begin{align} \label{eq:wrongcov}
\Gamma^{\rm naive}(v,iD) &= 
 1 + {i\Dslash_\perp\over 2 M} + {1\over M^2}\left[ -\frac18 (iD_\perp)^2 
- \frac12 i \Dslash_\perp iv\cdot D 
\right] 
\\
&\quad
+ {1\over M^3} \left[ \frac14 (iD_\perp)^2 iv\cdot D 
+ { i\Dslash_\perp \over 2} \left(
- {3\over 8} (iD_\perp)^2  
+ (iv\cdot D)^2 \right) 
\right] + \order(1/M^4) 
\,,  \nonumber
\end{align}
is not a solution to the invariance equation. 
The necessity for such additional field strength dependent terms can also be seen from the fact that
the right hand side of~\eqref{eq:wrongcov} would imply a transformation $\psi_v \to \psi_w 
= \Gamma^{\rm naive}(w,iD)^{-1} e^{iq\cdot x} \Gamma^{\rm naive}(v,iD) \psi_v$ 
that takes $\psi_v$ outside of the assumed representation 
space, with $\slash{v}\psi_v = \psi_v$. 
In the heavy fermion Lagrangian, the effects of these field-strength dependent 
terms appear first  at order  ${\mathcal O}(1/M^4)$, 
where omission of the final term in (\ref{eq:sol}) would lead to incorrect 
$1/M^4$ Lagrangian coefficient relations~\cite{nrqedm4}.%
\footnote{When building invariant fermion bilinears, the leading terms involve 
$iv\cdot D$ multiplying $1/M$ corrections appearing in $\Gamma(v,iD)$.   Since such 
terms are eliminated in going to canonical form, nontrivial effects of the $1/M^3$ corrections
to $\Gamma(v,iD)$ appear first at order $1/M^4$.  
}

Before closing this section, let us summarize the value of the
invariant operator method. Appendix~\ref{sec:allorders} shows that we
can find a suitable covariantization of $W({\cal B}, i\partial + Mv)$
that allows solution of the invariance equation for $\Gamma(v,iD)$ 
to any order in $1/M$. 
Hence this method proves the existence of a covariantized boost
operator and a non-zero, Lorentz invariant Lagrangian to arbitrary order.
We may proceed in either of two ways to construct invariant Lagrangians. 
Firstly, we may proceed as in (\ref{eq:Lfull}), where we construct manifestly invariant 
interactions through some fixed order in $1/M$; to achieve canonical form we must then 
perform field redefinitions.   
Alternatively, we may proceed as in (\ref{eq:abelian}) (or its generalization to arbitrary $v$), 
armed with the knowledge that a suitable boost generator as in (\ref{eq:redef}) can be reconstructed 
order by order.  

%%%%%%%%%%%%%%%%%%%%%%%%%%%%%%%%%%%%%%%%%%%%%%%%%%%%%%%%%%%%
%%%%%%%%%%%%%%%%%%%%%%%%%%%%%%%%%%%%%%%%%%%%%%%%%%%%%%%%%%%%
\section{Effective field theories for massless
particles \label{sec:massless}} \countertozero

Although our primary focus has been on the constraints imposed by Lorentz invariance
in heavy particle effective field theories, it is interesting to consider 
the applications of other Lorentz representations.   Recall that for physical states, 
representations of the Lorentz group fall into distinct classes, depending 
on the nature of $p^0$ and $p^2$.  For example, in our heavy particle applications we
considered the little group for $p^0> 0$ and $p^2=M^2 > 0$.   

Consider now the case $p^0>0$ and $p^2=0$.
This applies to the collinear sector of soft-collinear effective 
theory~\cite{Bauer:2000ew,Bauer:2000yr,Chay:2002vy,Beneke:2002ph,Hill:2002vw}.   
The little group in this case is isomorphic 
to $E(2)$, the euclidean group of rotations and translations in two dimensions. 
In analogy to the construction in Section~\ref{sec:little}, let us consider the little
group defined by the invariant vector $E\, n$, where $n^2=0$ and $E$ is a reference energy.  
In order to define the 
induced representation, let us also introduce a timelike unit vector $v$ with $v^2=1$.%
\footnote{
In applications to heavy quark processes the vector $v$ is naturally identified with the 
reference vector for the heavy quark field.
}
Given $n$ and $v$ we may define an additional lightlike vector, 
\be
\bar{n}^\mu = {1\over n\cdot v} \left( 2v^\mu - {n^\mu \over n\cdot v} \right) \,,
\ee
satisfying $\bar{n}^2=0$ and $n\cdot \bar{n}=2$. 
In this section ({\it only}) we define perpendicular components $p_\perp$ 
with respect to $n$ and $\bar{n}$.  We also define vectors $p_+$ and $p_-$ along the 
$n$ and $\bar{n}$ directions respectively, 
\begin{align}
p^\mu &\equiv {\bar{n}\cdot p \over 2} n^\mu + {n\cdot p\over 2} \bar{n}^\mu + p_\perp^\mu 
\equiv p_+^\mu + p_-^\mu + p_\perp^\mu \,. 
\end{align}
With this notation let us define a standard Lorentz transformation taking $En$ to $p$ 
as
\be \label{eq:lightlikeLp}
L(p) = L_{\bar{\cal S}}(p) L_{\cal B}(p) \,,
\ee
where $L_{\cal B}$ is a boost that takes $E n^\mu$ to $p_+^\mu$ and $L_{\bar{\cal S}}$ is a parabolic Lorentz transformation 
taking $p_+^\mu$ to $p^\mu$. 
They are given by
\begin{subequations}
\begin{align}
L_{\cal B}(p)^\mu_{\,\,\nu} & = g^\mu_{\,\,\nu} 
+ \frac12 \left( {\bar{n}\cdot p \over 2 E} - 1 \right) n^\mu \bar{n}_\nu 
+ \frac12 \left( {2 E \over \bar{n}\cdot p} - 1 \right) \bar{n}^\mu n_\nu \,, \\
L_{\bar{\cal S}}(p)^\mu_{\,\,\nu} & = g^\mu_{\,\,\nu}
+ {1\over \bar{n}\cdot p} \left( p_\perp^\mu \bar{n}_\nu - \bar{n}^\mu p_{\perp\nu} \right)
- {p_\perp^2\over 2 (\bar{n}\cdot p)^2 } \bar{n}^\mu \bar{n}_\nu \,. 
\end{align}
\end{subequations}
The choice (\ref{eq:lightlikeLp}) for $L(p)$
is convenient due to the resulting simplicity of $W(\Lambda,p)$.  
The space of physical states generated by $U[L(p)]|En^\mu,\sigma\rangle$ is sufficient to describe 
particles of a given helicity with non-vanishing $\bar{n}\cdot p$.     

It is straightforward to compute the little group element corresponding to arbitrary 
Lorentz transformations according to~\eqref{eq:WLp}.   
The six independent Lorentz transformations can be grouped into four classes.
First, there is the one-parameter group of 
rotations $\mathcal R$ that keep $n$ and $\bar n$ fixed.
Second, there is the two-parameter group of parabolic Lorentz transformations $\mathcal S$
that keep $n$ fixed but change $\bar n$. 
These two classes form the little group of $n$.
Third, there is the one-parameter 
group of boosts ${\mathcal B}$ in the $n$ direction that change $n$ and $\bar{n}$. 
Fourth, there is the two-parameter group of parabolic transformations $\bar {\mathcal S}$ 
that keep $\bar{n}$ fixed but change $n$.
In infinitesimal form these transformations are given by
\begin{subequations} \label{eq:lightlikeLT}
\begin{align}
{\cal R}(\theta)^\mu_{\,\,\nu} & = g^{\mu}_{\,\,\nu }+ \theta \epsilon^\mu_{\,\,\nu\rho\sigma} n^\rho \bar{n}^\sigma +\order(\theta^2) 
& 
 [ & n\to n, \bar n \to \bar n]    \,, \\
{\cal S}(\alpha)^\mu_{\,\,\nu} &  = g^\mu_{\,\,\nu} + \frac{ \alpha^\mu n_\nu - n^\mu \alpha_\nu }{2}  + \order(\alpha^2) \, 
& 
 [ & n\to n, \bar n \to \bar n +  \alpha]    \,, \\
{\cal B}(\eta)^\mu_{\,\,\nu} & = g^\mu_{\,\,\nu} + {\eta}\frac{ n^\mu \bar{n}_\nu - \bar{n}^\mu n_\nu}{2} 
+ \order(\eta^2) \,
& 
 [ & n\to (1+\eta)n, \bar n \to (1-\eta)\bar n]    \,, \\
\bar{\cal S}(\beta)^\mu_{\,\,\nu} & =  g^\mu_{\,\,\nu} +\frac{ \beta^\mu \bar{n}_\nu - \bar{n}^\mu \beta_\nu}{2} + \order(\beta^2)\,  
& 
 [ & n\to n + \beta, \bar n \to \bar n]    \,, 
\end{align}
\end{subequations}
where  $\alpha^\mu = \alpha_\perp^\mu$ and $\beta^\mu = \beta_\perp^\mu$.
Note that physical states must transform trivially under ${\mathcal S}$ to avoid continuous 
internal degrees of freedom and that 
little group elements can be parameterized as (e.g. see~\cite{Weinberg:1995mt})
\be\label{eq:littlelight}
W(\Lambda,p) = {\cal S}[\tilde{\alpha}(\Lambda,p)] {\cal R}[\tilde{\theta}(\Lambda,p)]  \,.
\ee
We find that the mapping \eqref{eq:WLp} with $L(p)$ chosen as in Eq.~\eqref{eq:lightlikeLp} 
takes the little group rotation ${\cal R}(\theta)$ into itself. 
Of the remaining three cases only the little group elements ${\cal S}(\alpha)$ 
have a non-trivial mapping
\begin{subequations} \label{eq:bar}
\begin{align}
\tilde{\theta}[{\cal R}(\theta),p] &  = \theta \,,  & \tilde{\alpha}[{\cal R}(\theta),p]^\mu & = 0 \, , \label{eq:bar1} \\
\tilde{\theta}[{\cal S}(\alpha),p] & = -{1\over 2( \bar{n}\cdot p)} \epsilon_{\mu\nu\rho\sigma} \alpha^\mu p_\perp^\nu n^\rho \bar{n}^\sigma \,,   	& \tilde{\alpha}[{\cal S}(\alpha),p]^\mu & = {E\over \bar{n}\cdot p} \alpha^\mu \, , \label{eq:bar2} \\
\tilde{\theta}[{\cal B}(\eta),p] & = 0 \,, & \tilde{\alpha}[{\cal B}(\eta),p]^\mu & = 0 \, , \label{eq:bar3} \\
\tilde{\theta}[\bar{\cal S}(\beta),p] & = 0 \,, &  \tilde{\alpha}[\bar{\cal S}(\beta),p]^\mu & = 0 \,. \label{eq:bar4} 
\end{align}
\end{subequations}
The result (\ref{eq:littlelight}) with little group parameters  \eqref{eq:bar}
defines the transformation law for particle states.
As in the timelike case, we postulate the field transformation law, 
\begin{equation} \label{eq:little_light_transform}
\phi_a(x) \to D[ W(\Lambda, iD_\mu ) ]_{ab} \phi_b(\Lambda^{-1}x) \,,
\end{equation}
where now $D(W)$ refers to a representation of the $E(2)$ little group. 

We focus on the representation appropriate to a massless spin-$1/2$ particle, 
\be\label{eq:lightD}
D[{\cal S}(\tilde{\alpha}){\cal R}(\tilde{\theta}) ]  = \exp[i\tilde{\theta}/2] \,, 
\ee
and embed this representation into a Dirac spinor representation $\psi_n$ of the Lorentz group. 
A trivial action of $\mathcal S$ on this field is equivalent to the constraint
\be \label{eq:project}
\slash{n} \psi_n = 0 \,. 
\ee
The transformation law,
\be\label{eq:lightlike}
\psi_n(x) \to \left( 1 + {i\over 4} \omega(\Lambda,iD)_{\mu\nu} \sigma^{\mu\nu}  \right) \psi_n(\Lambda^{-1}x) \,,
\ee
with $\omega_{\mu\nu}(\Lambda,iD)$ obtained from~\eqref{eq:bar} and $\psi_n$ satisfying (\ref{eq:project}),
reduces to (\ref{eq:lightD}).

Similar to the timelike case, we may investigate general conditions under which 
(\ref{eq:lightlike}) leads to a Lorentz invariant theory.   
We note that for terms in the fermion Lagrangian of the form
\be\label{eq:schematicn}
\bar{\psi}_n \bigg\{ \cdots n^\mu \dots \bar{n}^\mu \cdots D^\mu \cdots \gamma^\mu \cdots \bigg\} \psi_n \, ,
\ee 
we may recast invariance under (\ref{eq:lightlike}) 
as a collection of ``reparameterization'' transformations 
acting on $n$ and $\bar{n}$, cf Section~\ref{sec:reform}.   
In particular, invariance under rotations ${\cal R}(\theta)$ is ensured by writing a naively 
covariant Lagrangian in terms of the constrained field $\psi_n$, as in (\ref{eq:schematicn}).   
Transformations  $\bar{{\cal S}}(\beta)$, ${\cal S}(\alpha)$ and ${\cal B}(\eta)$ translate to 
the ``type-I'', ``type-II'' and ``type-III'' transformations considered in \cite{Manohar:2002fd}. 
A more detailed discussion of the lightlike case, involving 
a rigorous discussion of Lorentz invariance, and the inclusion of multiple momentum modes
and multiple gauge symmetries, is beyond the scope of the present paper and is left to future work. 

%%%%%%%%%%%%%%%%%%%%%%%%%%%%%%%%%%%%%%%%%%%%%%%%%%%%%%%%%%%%
%%%%%%%%%%%%%%%%%%%%%%%%%%%%%%%%%%%%%%%%%%%%%%%%%%%%%%%%%%%%
\section{Summary \label{sec:discuss}}
\countertozero

The usual procedure of implementing Lorentz invariance via 
finite dimensional representations of the Lorentz group is insufficient 
for application to heavy particle effective theories.  
We have adapted the formalism of induced representations for application to 
heavy particle field transformation laws.   
Returning to the questions posed in the Introduction, 
we see that the parameter $v$ enters as an arbitrary reference vector 
in the effective theory construction.
Rules identifiable with ``reparameterization invariance'' 
(\ref{eq:covrep}) are obtained by a rewriting of the transformation
law for generalized boosts, and  
the class of reparameterization transformations consistent with 
Lorentz and gauge invariance is identified
through a systematic solution of the invariance equation (\ref{eq:cons2}).
While an  explicit construction such as in (\ref{eq:HQET}) must map into this 
framework, 
it is not necessary to refer to a specific 
underlying ultraviolet completion, or to explicitly integrate out degrees of freedom
when deriving these transformation laws.  

Let us compare our formalism to previous work. 
A naive ansatz for implementing Lorentz invariance via 
reparameterization invariance breaks down for 
$\Gamma(v,iD)$ starting at order $1/M^3$, corresponding to 
new effects at order $1/M^4$ in the canonical Lagrangian.  
The transformation law defined by $W(\Lambda,iD)$ is corrected 
at order $1/M^4$.  
These subtleties were not treated 
in the classic work of Luke and Manohar~\cite{Luke:1992cs,Manohar:1997qy}, and the 
ansatz proposed there would lead to inconsistencies at the orders in $1/M$ specified above.   
Brambilla et al.~\cite{Brambilla:2003nt} recognized that Wilson-coefficient dependent 
corrections to $W(\Lambda)$ must be included when deriving an invariant Lagrangian 
in canonical form.  
However, there the constraints of Lorentz invariance are derived (through order $1/M^2$) 
at the level of canonically quantized charges, a procedure that becomes increasingly 
cumbersome at high orders in the $1/M$ expansion.   
In Section~\ref{sec:induced} 
we have used general properties of commutators of 
the $S$ matrix with conserved charges to 
derive constraints at the Lagrangian level that implement Lorentz invariance for
heavy particle effective theories in canonical form. 
In Section~\ref{sec:reparam} we have derived consistent reparameterization transformations that
allow solution to the invariance equation (\ref{eq:cons2}), and hence the construction of
manifestly invariant Lagrangians to arbitrary order.  

We demonstrated the application of our formalism in the case of NRQED (i.e., the
parity and time-reversal symmetric theory of a heavy spin-$1/2$
particle coupled to an abelian gauge field). 
At a practical level, the main results for building heavy fermion Lagrangians 
are contained in~\eqref{eq:redef}, or for the invariant operator method, 
in~\eqref{eq:gammaexpand} and \eqref{eq:gammaint}.
The NRQED Lagrangian is computed at $\order(1/M^4)$ in \cite{nrqedm4}. 

We note that a choice must be made between a canonical form of the Lagrangian with somewhat complicated 
boost generator, versus a simpler form of the boost generator with non-canonical Lagrangian.   
In practical computations, it is typically easier to choose the former approach.  
We remark that a regularization scheme that breaks Lorentz symmetry must be accompanied by 
counterterms that reinstate the symmetry.%
\footnote{
We thank T. Becher for a discussion on this point.
}  
Renormalization of the Lagrangian in canonical form should be defined in such a way that
non-canonical terms are not generated. 

The heavy particle limit considered here assumes a single large mass scale.  
Interesting complications can arise when this is not the case, 
e.g. in the phenomenology of heavy baryons in low-energy processes involving pions, 
$\Delta$ excitations and electroweak gauge 
interactions.   
Numerically large coefficients appearing in the $m_\pi/m_N$ expansion limit the usefulness of 
the heavy particle expansion unless certain formally suppressed terms are ``resummed'', introducing
nontrivial power counting and renormalization
issues~\cite{Gasser:1987rb,Jenkins:1990jv,Becher:1999he,Fuchs:2003qc,Pascalutsa:2005nd,Hill:2009ek}. 
While it may be possible to embed a given heavy particle theory into a larger structure, 
this does not lessen the importance of understanding Lorentz invariance in the low energy limit.%
\footnote{This may be viewed in analogy to embedding nonlinear sigma models into linear sigma models
with extra degrees of freedom.}

The formalism presented here can be applied to straightforwardly
construct heavy particle Lagrangians of arbitrary spin. 
It can also be easily extended to include multiple heavy particle fields, 
and other relativistic degrees of freedom
beyond the abelian gauge fields considered here.   
As described in Section~\ref{sec:massless} 
the extension to effective field theories for massless particles
involves induced representations for little group isomorphic to 
$E(2)$, the euclidean transformations in two dimensions. 
A rigorous analysis along these lines may help clarify several outstanding issues 
in SCET, ranging from the appearance of new momentum modes, to the interplay 
of ultraviolet regulators and factorization~\cite{Becher:2003qh,Becher:2010tm,Chiu:2011qc}. 
It may be interesting to investigate the application of the little group 
corresponding to a spacelike reference vector, $s^2=-1$ (cf. our $v^2=1$ and $n^2=0$
cases), and to explore embeddings into nonlinear realizations with fictitious 
Goldstone fields~\cite{Volkov:1973vd}. 

\vskip 0.2in
\noindent
{\bf Acknowledgements}
\vskip 0.1in
\noindent
We thank G. Lee and G. Paz for collaboration on explicit computations with the $1/M^4$ NRQED Lagrangian
that helped reveal the subtleties in reparameterization invariance at this order, and for comments on 
the manuscript.  Work supported by NSF Grant 0855039 and DOE Grant DE-FG02-90ER-40560. 

%%%%%%%%%%%%%%%%%%%%%%%%%%%%%%%%%%%%%%%%%%%%%%%%%%%%%%%%%%%%
%%%%%%%%%%%%%%%%%%%%%%%%%%%%%%%%%%%%%%%%%%%%%%%%%%%%%%%%%%%%
%%%%%%%%%%%%%%%%%%%%%%%%%%%%%%%%%%%%%%%%%%%%%%%%%%%%%%%%%%%%
%%%%%%%%%%%%%%%%%%%%%%%%%%%%%%%%%%%%%%%%%%%%%%%%%%%%%%%%%%%%
\begin{appendix}

%%%%%%%%%%%%%%%%%%%%%%%%%%%%%%%%%%%%%%%%%%%%%%%%%%%%%%%%%%%%
%%%%%%%%%%%%%%%%%%%%%%%%%%%%%%%%%%%%%%%%%%%%%%%%%%%%%%%%%%%%
\section{Extension to arbitrary spin and self-conjugate fields\label{sec:covariant}} 
\countertozero

Although the explicit results in this paper are focused on spin-$1/2$ fields transforming under
an abelian (i.e. complex) gauge group,
the formalism extends straightforwardly 
to fields of arbitrary spin or to self-conjugate fields.
In section~\ref{sec:higherspin} we describe the formalism for embedding arbitrary
spin representations within products of Dirac spinor and Lorentz vector representations 
of the Lorentz group. For a related discussion see e.g.~\cite{Falk:1991nq}. 
Section~\ref{sec:self} describes the constraints imposed on 
the effective theory deriving from self-conjugate fields. 
For a related discussion see e.g.~\cite{Hill:2011be}.

%%%%%%%%%%%%%%%%%%%%%%%%%%%%%%%%%%%%%%%%%%%%%%%%%%%%%%%%%%%%
\subsection{Higher spin representations \label{sec:higherspin}}

Irreducible 
higher spin representations 
can be built 
using products of the Dirac spinor and vector
representations 
\be \label{eq:trans2}
\psi_v \to 
\Lambda_\frac12 \psi_v  \,, 
\quad 
Z^\alpha_v \to  
 \Lambda^\alpha_{\,\,\beta} Z^\beta_v \, ,
\ee
where $\Lambda=D(W)$ is a little group element as  in Section~\ref{sec:little}, i.e., $\Lambda v = v$. The corresponding generators for these two representations are given by
\begin{equation} \label{eq:Jmunu}
{\mathcal J^{\alpha\beta}_\frac{1}{2}} = \frac{1}{2} \sigma^{\alpha\beta} = \frac{i}{4} [\gamma^\alpha, \gamma^\beta] \,,
\qquad 
({\mathcal J^{\alpha\beta}})_{\mu\nu} = i (g^{\alpha}_{\,\,\mu} g^{\beta}_{\,\,\nu}- g^{\beta}_{\,\,\mu} g^{\alpha}_{\,\,\nu}).
\end{equation}
We enforce a maximal set of constraints to isolate the appropriate irreducible 
representation. 

\begin{description}
\item[Integer spin:]
For integer spin $s=n$, consider the totally symmetric and traceless
tensor $Z_v^{\mu_1... \mu_n}$, which has $(n+1)^2$ degrees of freedom. 
Imposing
\be\label{eq:vconstraint}
v_{\mu_1} Z_v^{\mu_1 ... \mu_n} =0 
\ee
yields $n^2$ additional constraints, leaving us with $2n+1=2s+1$ degrees of freedom as desired. 
Under Lorentz transformations this field transforms as
\begin{equation}
Z^{\mu_1... \mu_n}_v \to  
 \Lambda^{\mu_1}_{\,\,\nu_1}...  \Lambda^{\mu_n}_{\,\,\nu_n} Z^{\nu_1... \nu_n}_v \, .
\end{equation}
Using $\Lambda^T g \Lambda = g$ and $\Lambda v = v$, 
it is easy to see that symmetry, tracelessness and the constraint \eqref{eq:vconstraint}  are preserved by this transformation.
\item[Half-integer spin:]
For half-integer spin $s=n+1/2$, consider the 
spinor-tensor $\psi_v^{\mu_1, \mu_2, ..., \mu_n}$, which is totally 
symmetric in the indices $\mu_1...\mu_n$ and therefore has $2(n+1)(n+2)(n+3)/3$  degrees of freedom.  
We impose the constraints~%
\footnote{Note that the second constraint implies $g_{\mu\nu}\psi_v^{\mu \nu \mu_3 ...\mu_n} =0$ 
and, furthermore, is equivalent to imposing $v_{\mu_1} \psi_v^{\mu_1...\mu_n} =0$ 
and $\epsilon_{\nu\alpha\beta\mu_1}v^{\nu} \sigma^{\alpha\beta} \psi_v^{\mu_1...\mu_n} =0$ }
\be \label{eq:vconstrain12}
\slash{v} \psi_v^{\mu_1... \mu_n} = \psi_v^{\mu_1... \mu_n} \,, \quad  \gamma_{\mu_1}
\psi_v^{\mu_1...\mu_n} =0.
\ee
The second constraint yields $n(n+1)(n+5)/3$ equations, 
while the first projects  a four-component spinor onto a two-dimensional 
subspace, reducing the degrees of freedom by $1/2$. 
In total $2(n+1)=2s+1$ degrees of freedom remain. 
Under Lorentz transformations this field transforms as
\begin{equation}
\psi^{\mu_1... \mu_n}_v \to  
 \Lambda^{\mu_1}_{\,\,\nu_1}...  \Lambda^{\mu_n}_{\,\,\nu_n} \Lambda_{\frac12} \psi^{\nu_1... \nu_n}_v \,.
\end{equation}
This is symmetric in  $\mu_1...\mu_n$.  
That equations \eqref{eq:vconstrain12} are preserved follows 
immediately from $\Lambda v = v$ and $\Lambda_{\frac12}^{-1} 
\gamma^\mu \Lambda_{\frac12} =  \Lambda^{\mu}_{\,\,\nu} \gamma^\nu$.
\end{description}

%%%%%%%%%%%%%%%%%%%%%%%%%%%%%%%%%%%%%%%%%%%%%%%%%%%%%%%%%%%%
\subsection{Self-conjugate fields \label{sec:self}}
The self-conjugacy of $SU(2)$ implies that for any field $\phi(x)$ transforming 
as in (\ref{eq:standardgen}) or (\ref{eq:infin})
with the plus sign,
the field
\be \phi^c(x) = S \phi^*(x) \,, 
\ee
transforms as in  (\ref{eq:standardgen}) or (\ref{eq:infin})
with the minus sign. 
Here $S$ is the $(2s+1)\times (2s+1)$ similarity transformation 
for the spin-$s$ representation of
$SU(2)$, such that $(-\Sigma^i)^* = S \Sigma^i S^{-1}$.  
In covariant language, this translates to the simultaneous transformations
\be \label{eq:parity1}
\phi_v(x) \to \phi_v^c(x)\,, \quad 
v^\mu \to - v^\mu \,. 
\ee
In terms of the irreducible representations constructed in Section~\ref{sec:higherspin}, 
the field transformation in (\ref{eq:parity1}) reads%
\footnote{We here choose a basis such that $S=1$ for vectors.}
\be \label{eq:parity2} 
Z^{\mu_1... \mu_s}_v \to (Z_v^c)^{\mu_1... \mu_s} = (Z_v^{\mu_1 ... \mu_s})^*
\,, \quad 
\psi_v^{\mu_1... \mu_s} \to (\psi_v^c)^{\mu_1... \mu_s} = \mathcal{C} (\psi_v^{\mu_1 ... \mu_s})^*\, , 
\ee
for integer spin and half-integer spin fields, respectively. 
The charge conjugation matrix 
$\mathcal{C}$ acts on the spinor index of $\psi_v$.  
It is symmetric and  unitary, and 
obeys $\mathcal{C}^\dagger \gamma^\mu \mathcal{C} = - \gamma^{\mu*}$. 
The parity (\ref{eq:parity1}) arises if the effective theory is describing 
a full theory of a self-conjugate field
(necessarily transforming in a real representation of a gauge group). 
For example, the effective theory field for a real scalar $\varphi=\varphi^*$ 
can be obtained via
\be 
\varphi(x) = e^{-iMv\cdot x} \varphi_v(x)/\sqrt{M} = e^{iMv\cdot x} \varphi_v^*(x)/\sqrt{M} = \varphi^*(x) \,. 
\ee
Similarly, the effective theory for a Majorana fermion represented by a Dirac spinor $\psi_M=\psi_M^c$ 
can be obtained via 
\be
\psi_M = \sqrt{2}e^{-iMv\cdot x} (h_v + H_v) = \sqrt{2}e^{iMv\cdot x} ( h_v^c + H_v^c ) = \psi_M^c \,, 
\ee
where $\slash v h_v= h_v$ and $\slash v H_v = - H_v$. 

It follows from (\ref{eq:parity1}) that the allowed operators
$\bar{ \phi}_v \mathcal{O}(v) \phi_v$ in the Lagrangian representing a self-conjugate 
field
can be chosen such that 
\begin{equation} \label{eq:selfO}
\mathcal{O}(v) = {\cal C} \mathcal{O}(-v)^* {\cal C}^\dagger.
\end{equation} 
Since we are often interested in constructing the Lagrangian in canonical form, 
i.e., without higher $iv\cdot D$ derivatives acting on $\phi_v$, it is important to 
ask whether this condition is preserved by the requisite field redefinitions. 
By a similar reasoning to above, operators of the form 
$\bar{\phi}_v [ iv \cdot D X(v) + X^\dagger(v) iv \cdot D ]\phi_v$ 
appearing in
the Lagrangian must be such that $X(v) = {\cal C} X(-v)^* {\cal C}^\dagger$. 
Hence field redefinitions of the form $\phi_v \to \left[1- X(v)\right] \phi_v$ 
achieve canonical form of the Lagrangian while preserving \eqref{eq:selfO}. 

%%%%%%%%%%%%%%%%%%%%%%%%%%%%%%%%%%%%%%%%%%%%%%%%%%%%%%%%%%%%
%%%%%%%%%%%%%%%%%%%%%%%%%%%%%%%%%%%%%%%%%%%%%%%%%%%%%%%%%%%%
\section{Solution to the invariance equation \label{sec:allorders}} 
\countertozero

Section~\ref{sec:gammasol} describes the solution of the invariance 
equation (\ref{eq:cons2}) for the function $\Gamma(v,iD)$ in the free theory.
The solution in the interacting theory is not simply obtained from the free one by replacing $\partial$ with $D$.
Here we present a method of solution that is valid to any order in $1/M$. Since we use $\Gamma(v,iD)$ to 
construct the invariant Lagrangian, the existence of a solution for $\Gamma(v,iD)$ proves that a non-zero Lagrangian
exists at any order in $1/M$. First, we will  construct the general solution  in section~\ref{sec:generalGammasol}  and then 
explicitly apply this construction to the spin 1/2 theory up to order $1/M^3$ in section~\ref{sec:explicitGammasolution}.

%%%%%%%%%%%%%%%%%%%%%%%%%%%%%%%%%%%%%%%%%%%%%%%%%%%%%%%%%%%%
\subsection{Series solution for $\Gamma$} \label{sec:generalGammasol}

Recall the equation (\ref{eq:cons2}) for $\Gamma$ required to build explicitly invariant 
operators, 
\be
\Gamma(v+q/M, iD-q)   {\cal B}^{-1}  {W}({\cal B}, iD + M v) = \Gamma(v, i D) \,,
\ee
where to first order in $q$ we have 
${\cal B}^{-1} v = v + q/M$. Let us expand in orders of $1/M$ and define
\begin{subequations}
\begin{align}
X \equiv {\cal B}^{-1} W &  = 1 + q^\mu X_\mu 
= 1+ q^\mu \left[ {1\over M } X^{(1)}_\mu + {1\over M^2} X^{(2)}_\mu + \dots \right] \,, \label{eq:defofX} \\
\Gamma & = 1 + {1\over M} \Gamma^{(1)} + {1\over M^2} \Gamma^{(2)} + \dots \,.
\end{align}
\end{subequations}
We note that the variation in $\Gamma$ arises from the variations in 
$v$ and in $iD$, 
\be
\delta \Gamma = \Gamma(v+q/M,iD-q) - \Gamma(v,iD) 
= q^\mu \left( - {\partial \over \partial iD^\mu} \Gamma + {1\over M} {\partial \over \partial v^\mu} \Gamma \right) \,. 
\ee
Equating orders in $1/M$, we find
\begin{align}\label{eq:Gamma}
{\partial\over \partial i D^\mu} \Gamma^{(n)} 
&= {\partial \over \partial v^\mu} \Gamma^{(n-1)} + \Gamma^{(n-1)} X^{(1)}_\mu + \Gamma^{(n-2)} X^{(2)}_\mu 
+ \dots  + \Gamma^{(0)} X^{(n)}_\mu 
\equiv Y^{(n)}_\mu \,, 
\end{align}
where we define $\Gamma^{(0)}=1$.
Note that Eq.~\eqref{eq:Gamma} is understood to be contracted with $q^\mu$ so that  pieces proportional to $v^\mu$ should be dropped. 
We can solve this equation for $\Gamma^{(n)}$ obtaining
\begin{align}\label{eq:Gammasol}
\Gamma^{(n)} 
&= \sum_{m=1}^n {(-1)^{m-1}\over m !} iD_\perp^{\mu_1} iD_\perp^{\mu_2} \dots i D_\perp^{\mu_m} 
{\partial \over \partial i D^{\mu_1} }
{\partial \over \partial i D^{\mu_2} }
\dots
{\partial \over \partial i D^{\mu_{m-1}} }
Y^{(n)}_{\mu_m}   
 \nl
 &= iD_\perp^\mu Y^{(n)}_\mu - {1\over 2 !} iD_\perp^\mu iD_\perp^\nu 
 {\partial \over \partial iD^\mu} Y^{(n)}_\nu 
 + \dots 
\, ,
\end{align}
provided that  at each order, the $Y^{(n)}$ derived from the already determined 
$\Gamma^{(1)}\,,\dots\,,\Gamma^{(n-1)}$ satisfy%
\footnote{
This is the analog of  $\vec{\nabla}\times \vec{E} = \vec{0}$ for the existence of 
a solution $\phi$ of $\vec{\nabla} \phi = \vec{E}$ in electrostatics. 
}
\be\label{eq:Y}
{\partial\over \partial i D^{[\nu} } Y^{(n)}_{\mu]} =  0 \,,
\ee
where $A^{[\mu}B^{\nu]} = (A^\mu B^\nu - A^\nu B^\mu)/2$ denotes antisymmetrization.   
Using the definition of $Y^{(n)}$ we can show that this imposes constraints on $X^{(n)}$, for $n\ge 2$,
\begin{align}\label{eq:Xconstraint}
{\partial \over \partial i D^{[\nu} } X^{(n)}_{\mu]} 
&= - {\partial \over \partial v^{[\mu}} X^{(n-1)}_{\nu]}  
+ X^{(n-1)}_{[\mu} X^{(1)}_{\nu]} 
+ X^{(n-2)}_{[\mu} X^{(2)}_{\nu]} 
+ \dots 
+ X^{(1)}_{[\mu} X^{(n-1)}_{\nu]} \equiv Z^{(n)}_{\mu\nu} \,.
\end{align}
For Eq.~\eqref{eq:Xconstraint} to have a solution, a consistency condition on $Z^{(n)}_{\mu\nu}$ requires that%
\footnote{
This is the analog of $\vec{\nabla}\cdot\vec{B} = 0$ for the existence of a solution 
$\vec{A}$  of $\vec{\nabla}\times \vec{A} = \vec{B}$ in magnetostatics.
}
\be\label{eq:Zconsistency}
0 = v_\sigma \epsilon^{\mu\nu\rho\sigma} {\partial \over \partial i D^\rho} Z^{(n)}_{\mu\nu} \,. 
\ee
We can show by induction that Eq.~\eqref{eq:Xconstraint} can be solved
at each order.  
Since $X^{(1)}$ is dimensionless, it cannot depend on $iD$;  
hence $Z^{(2)}$ from (\ref{eq:Xconstraint}) is also independent of $iD$ 
and solves (\ref{eq:Zconsistency}). 
Now assume that we have constructed solutions $X^{(n)}$
to Eq.~\eqref{eq:Xconstraint} for $n=1,...,N-1$
(necessarily obeying the constraint~\eqref{eq:Zconsistency}). 
Application of the Jacobi identity 
shows that the constraint~\eqref{eq:Zconsistency} is then obeyed for $n=N$ and a
solution to Eq.~\eqref{eq:Xconstraint} can be found for $n=N$.

Let us find a solution to
Eq.~\eqref{eq:Xconstraint} that reduces to a given $X_{\rm free}$ for the 
non-interacting theory (e.g., $X_{\rm free} = {\cal B}^{-1} W$ from (\ref{eq:Wfree}) ).  
First, note that the existence of the free case
solution given in~\eqref{eq:gammafree} implies that the $X^{(n)}$
defined in the free case  from~\eqref{eq:Wfree} must obey the
constraint~\eqref{eq:Xconstraint}. Let us define naively
covariantized quantities $\hat X^{(n)} = X^{(n)}_{\text{free}}\Big|_{\partial \to D}$,
with a definite ordering prescription, e.g. as in (\ref{eq:wrongcov}),
and define $\hat Z^{(n)}$ by
\begin{equation}
\hat Z^{(n)}_{\mu\nu} \equiv {\partial \over \partial i D^{[\nu} }
\hat X^{(n)}_{\mu]} \,.
\end{equation}
A straightforward calculation then shows that~\eqref{eq:Xconstraint} is solved by
\begin{align} \label{eq:Xnsol}
X^{(n)}_{\mu} &= \hat X^{(n)}_\mu 
+ 
2 \sum_{m=1}^{n-1} {(-1)^{m}\over (m+1)!} iD^{\nu_1}_\perp \cdots iD^{\nu_m}_\perp
{\partial \over \partial iD^{\nu_1} } 
\cdots 
{\partial \over \partial iD^{\nu_{m-1}} } 
\left(Z^{(n)}_{\nu_m \mu}  - \hat{Z}^{(n)}_{\nu_m \mu} \right) 
\,.
\end{align}
In the free case we have $Z^{(n)} = \hat Z^{(n)}$ and  $X^{(n)}$ reduces to the free case solution.
Having found a suitable $X^{(n)}$ satisfying (\ref{eq:Xconstraint}) 
we may then proceed to build $\Gamma^{(n)}$ satisfying (\ref{eq:Gamma}), 
and hence $\Gamma$ satisfying (\ref{eq:cons2}). 

Note that $Z^{(n)}$ has mass dimension $n-2$ so that $n=4$
is the first order at which field strength dependent terms can cause  
$Z^{(n)} \neq \hat Z^{(n)}$.  
Correspondingly, our choice (\ref{eq:Xnsol}) ensures that 
field-strength dependent corrections 
to $X^{(n)}- \hat{X}^{(n)}$ can first appear at order $n=4$. 
This can be explicitly seen in the solution for the spin 1/2 theory in  the next section.

%%%%%%%%%%%%%%%%%%%%%%%%%%%%%%%%%%%%%%%%%%%%%%%%%%%%%%%%%%%%
\subsection{Explicit solution for $\Gamma$ in the spin 1/2 theory} \label{sec:explicitGammasolution}

To illustrate, let us calculate $\Gamma$ for the spin 1/2 theory.
 Consider the free solution (\ref{eq:Wfree}), 
\begin{multline}
{X}_\mu(v,i\partial) = \frac{1}{2M} \gamma^\perp_\mu + 
\frac{1}{4M^2} \sigma^\perp_{\mu\nu} \partial^\nu 
\bigg[ 1 - {iv\cdot \partial\over M} + {1\over M^2}\left( (iv\cdot\partial)^2 -\frac14 (i\partial_\perp)^2 \right)
+\dots
\bigg] \,, 
\end{multline}
and the arbitrary covariantization, 
\begin{multline}
\hat{X}_\mu(v,iD) = \frac{1}{2M} \gamma^\perp_\mu + 
\frac{1}{4M^2} \sigma^\perp_{\mu\nu} D^\nu 
\bigg[ 1 - {iv\cdot D\over M} + {1\over M^2}\left( (iv\cdot D)^2 -\frac14 (i D_\perp)^2 \right)
+ \dots 
\bigg] \,. 
\end{multline}
A corresponding solution for $\Gamma$ in the free theory is displayed in (\ref{eq:gammafree}). Now let us 
follow the construction of the previous section order by order.
\begin{description}
%%%%%%%%%%%%%%%%%%%%%%%%
\item[Order $1/M$:]
First, we determine,
\be
Y^{(1)}_{\mu} 
= X^{(1)}_\mu = \hat{X}^{(1)}_\mu 
= {\gamma^\perp_\mu\over 2}  \,.
\ee
This function clearly satisfies
Eq.~\eqref{eq:Y}
so that we may solve for
\be
\Gamma^{(1)} 
= \frac12 i\Dslash_\perp \,.
\ee
%%%%%%%%%%%%%%%%%%%%%%%%
\item[Order $1/M^2$:]
Continuing to the next order, we evaluate 
\begin{subequations}
\begin{align}
Z^{(2)}_{\mu\nu} 
& =  
-{i\over 4} \sigma^\perp_{\mu\nu} = \hat{Z}^{(2)}_{\mu\nu} \,, \\
X^{(2)}_{\mu} & = \frac14 \sigma^\perp_{\mu\nu} D^\nu
 = \hat X^{(2)}_{\mu}  \,, \\
 Y^{(2)}_\mu 
 & = -\frac12 \gamma^\perp_\mu iv\cdot D - \frac14 i D^\perp_\mu \,. 
\end{align}
\end{subequations}
Solving for $\Gamma^{(2)}$ yields 
\be
\Gamma^{(2)} 
= 
- \frac18 (iD_\perp)^2 -\frac12 i\Dslash_\perp iv\cdot D   \,.
\ee
%%%%%%%%%%%%%%%%%%%%%%%%
\item[Order $1/M^3$:]
At the next order, we find 
\begin{subequations}
\begin{align}
Z^{(3)}_{\mu\nu} & = \frac{i}{4} \sigma^\perp_{\mu\nu} iv\cdot D  = \hat{Z}^{(3)}_{\mu\nu} \,, \\
X^{(3)}_{\mu} & = -\frac14 \sigma^\perp_{\mu\nu} D^\nu iv\cdot D=
\hat X^{(3)}_{\mu} \,, \\ 
Y^{(3)}_\mu &= \frac12 \gamma^\perp_\mu (iv\cdot D)^2 + \frac38 iD^\perp_\mu iv\cdot D
+ \frac18 iv\cdot D iD^\perp_\mu 
-\frac12 i\Dslash_\perp iD^\perp_\mu 
\\ 
& \hspace{6cm}
 -\frac{1}{16} (iD_\perp)^2 \gamma^\perp_\mu  
+ \frac18 i\Dslash_\perp \sigma^\perp_{\mu\nu}D^\nu  \,. \nonumber
\end{align}
\end{subequations}
After some manipulations, the resulting $\Gamma^{(3)}$ is 
\begin{multline}
\Gamma^{(3)} =  \frac14 (iD_\perp)^2 iv\cdot D 
+ {i\Dslash_{\perp}\over 2} 
\left[- \frac{3}{8} i\Dslash_\perp (iD_\perp)^2 
+  (iv\cdot D)^2 
\right] 
- {g\over 8} v^\alpha G_{\alpha\beta} D_\perp^\beta 
-{g\over 16} \sigma^\perp_{\alpha\beta} G^{\alpha\beta} i\Dslash_\perp
\\
+ {g\over 8} \bigg[ 
 i \gamma_\perp^\beta \sigma_\perp^{\mu\alpha} [D_\mu, G_{\beta\alpha}] 
- v^\alpha [D_\perp^\mu, G_{\alpha\mu}] 
- [D_\perp^\mu, G^\perp_{\mu\beta}]\gamma_\perp^\beta 
\bigg] \, .
\end{multline}
%%%%%%%%%%%%%%%%%%%%%%%%
\item[Order $1/M^4$:]
Continuing to higher order we find 
\begin{subequations}
\begin{align}
Z^{(4)}_{\mu\nu} &= \hat{Z}^{(4)}_{\mu\nu} 
+ {g\over 32}\left(  -i G^\perp_{\mu\nu} 
+ \sigma^\perp_{\mu\sigma} G^{\perp\sigma}_{\nu}
 - \sigma^\perp_{\nu\sigma} G^{\perp\sigma}_{\mu} \right)\,, \\
X^{(4)}_\mu 
& = \sigma^\perp_{\mu\nu} D^\nu \left[ \frac14 (iv\cdot D)^2 - {1\over 16} (iD_\perp)^2 \right]
+ {g\over 32} iD_\perp^\nu \left(  -iG^\perp_{\mu\nu} 
+ \sigma^\perp_{\mu\sigma} G^{\perp\sigma}_{\nu}
 - \sigma^\perp_{\nu\sigma} G^{\perp\sigma}_{\mu} \right) \,. 
\end{align}
\end{subequations}
Note that $X^{(4)}_\mu$ differs from the trial solution $\hat{X}^{(4)}_\mu$.
We may continue in this manner to construct $Y^{(4)}_\mu$ and $\Gamma^{(4)}$.
\end{description}

\end{appendix}

%%%%%%%%%%%%%%%%%%%%%%%%%%%%%%%%%%%%%%%%%%%%%%%%%%%%%%%%%%%%
%%%%%%%%%%%%%%%%%%%%%%%%%%%%%%%%%%%%%%%%%%%%%%%%%%%%%%%%%%%%
%%%%%%%%%%%%%%%%%%%%%%%%%%%%%%%%%%%%%%%%%%%%%%%%%%%%%%%%%%%%
%%%%%%%%%%%%%%%%%%%%%%%%%%%%%%%%%%%%%%%%%%%%%%%%%%%%%%%%%%%%
\bibliographystyle{unsrtnatbib}

%%%%%%%%%%%%%%%%%%%%%%%%%%%%%%%%%%%%%%%%%%%%%%%%%%%%%%%%%%%%

\end{document}